\documentclass[preprint,prb,aps,endfloats]{revtex4}
\usepackage{dcolumn}
\usepackage[dvips]{graphicx}
\newcolumntype{.}{D{.}{.}{1}}
\begin{document}
\title{Mesophases in Nearly 2D Room-Temperature Ionic Liquids}
\author{N. Manini, M. Cesaratto,}
\affiliation{European Theoretical Spectroscopy Facility (ETSF)}
\affiliation{Dipartimento di Fisica, Universit{\`a} di Milano, Via Celoria 16, 20133 Milan, Italy} 
\author{M. G. Del P{\'o}polo, and P. Ballone}
\affiliation{Atomistic Simulation Centre, Queen's University Belfast,
Belfast BT7 1NN, UK}
\begin{abstract}
Computer simulations of (i) a [C$_{12}$mim][Tf$_2$N] film of nanometric thickness squeezed at kbar pressure
by a piecewise parabolic confining potential reveal a mesoscopic {\it in-plane} density and composition modulation reminiscent
of mesophases seen in 3D samples of the same room-temperature ionic liquid (RTIL). Near 2D confinement, enforced by 
a high normal load, and relatively long aliphatic chains are strictly required for the mesophase formation, as confirmed
by computations for two related systems made of (ii) the same [C$_{12}$mim][Tf$_2$N] adsorbed at a neutral solid surface,
and of (iii) a shorter-chain RTIL ([C$_{4}$mim][Tf$_2$N]) trapped into the potential well of part i. 
No in-plane modulation is seen for ii and iii. In case ii, the optimal arrangement of charge and neutral tails is 
achieved by layering parallel to the surface, while in case iii weaker dispersion and packing interactions are unable 
to bring aliphatic tails together into mesoscopic islands, against overwhelming entropy and Coulomb forces.
The onset of in-plane mesophases could greatly affect the properties of long-chain RTIL used as lubricants.
\end{abstract}
\date{\today}
\maketitle
\section{Introduction}
\label{intro}

Systems made of alkane-substituted imidazolium cations ([C$_n$mim]$^+$) paired to a variety of anions (X$^-$) such Cl$^-$, 
PF$_6^-$, BF$_4^-$, etc., are the most representative examples of room-temperature ionic liquids (RTIL)\cite{wasser}, a vast class
of organic ionic compounds whose melting temperature falls below $100$ $^{\circ}$C. Low volatility, low flammability,
relatively high thermal and electrochemical stability are among the desirable properties shared by
several RTIL,\cite{welton} making these compounds attractive candidates for applications\cite{green} as solvents, in catalysis, in 
electrochemistry and, last but not least, as lubricants \cite{lubricant}. 

From a more fundamental point of view, the interest is enhanced by the surprising variety of properties and behaviours that RTIL 
display, making them important prototypes of molecular Coulomb liquids\cite{popolo}. The spontaneous formation of mesoscopic structures 
(mesophases) in RTIL, in particular, first predicted by computer simulations\cite{meso1, meso2, meso3, meso4}, and later confirmed by
experiments\cite{expemeso}, is among the most fascinating phenomena discovered in these systems.

\begin{figure}
\includegraphics[width=6cm,angle=0]{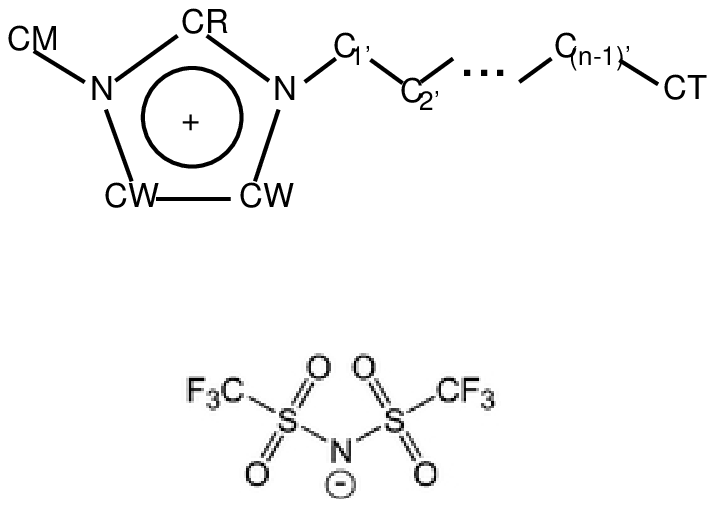}
\caption{Schematic structure of the [C$_n$mim][Tf$_2$N] ions considered in the simulations.
}
\label{scheme}
\end{figure}

Mesophases arise in RTIL's from the interplay of strong Coulomb interactions, weaker but still important dispersion forces, 
and steric effects.  Coulomb interactions, in particular, tend to concentrate the most ionic portions of the [C$_n$mim][X] molecules, 
represented by the positively charged imidazolium ring, and by the anion,
see Fig.~\ref{scheme}.
The neutral C$_n$H$_{2n+1}$ alkane tails, pushed out of these charged regions, 
group together under the effect of packing and van der Waals attraction.
Since the alkane chains cannot be physically separated from the cationic imidazolium ring to which they are connected by a covalent bond, a mesoscopic 
density and charge modulation is established, giving rise to the optimal separation of ionic moieties and alkane tails that is compatible with the 
molecular topology.
This picture is confirmed by the dependence of the mesophase stability on the length of the alkane tail attached to the imidazolium cation. 
Mesophases are observed only for RTIL's of sufficiently long tail ([C$_n$mim][X] with $n \gtrsim 6$), whose dispersion and steric interactions
are weaker but nevertheless comparable to Coulomb forces. 

Further insight on the origin and properties of mesophases could be gained by investigating systems of reduced dimensionality, since 
the interactions giving rise to the mesoscopic modulation are affected to a different degree by a change in the number of dimensions. 
The net effect of dimensional changes on the stability of mesophases, however, is difficult to predict a priori.
Reducing dimensionality reduces the strength of Coulomb interactions, thus reducing the primary driving force for the mesophase formation.
On the other hand, decreasing dimensionality limits screening and enhances fluctuations, possibly lending stability to mesophases.

We apply molecular dynamic simulations to investigate the stability and role of mesophases for long alkane tail RTIL in quasi-2D systems. 
Three different cases are considered.
In the first case (i), [C$_{12}$mim][Tf$_2$N] ions are trapped at $\sim 10^2$ MPa (kbar) pressure in between two parallel solid surfaces,
represented by a planar potential piecewise parabolic in the $z$ direction.
In the second case (ii), a nanometric [C$_{12}$mim][Tf$_2$N] film is adsorbed at the solid-vacuum interface, 
represented by an asymmetric, $z$-dependent potential arising from a flat and rigid surface, again parallel to the $xy$ plane. 
In the third case (iii), a shorter-tail RTIL ([C$_4$mim][Tf$_2$N]) is confined by the same potential of case i. The schematic structure of all these
molecules is shown in Fig.~\ref{scheme}.

Mesophases are apparent in simulation snapshots for case i, corresponding to a thin [C$_{12}$mim][Tf$_2$N] film trapped in 
between two planar and rigid surfaces, displaying nanometric features qualitatively similar to those seen in 3D systems. 
The degree of ordering and its characteristic length are quantified by computing the in-plane 2D structure factors 
[$S_{\alpha \beta}(q)$, $\alpha, \beta= +$, $-$] for the ionic moieties, i.e., the imidazolium ring cation and the [Tf$_2$N]$^-$ anion. 
The $S_{\alpha \beta}(q)$'s from the simulation display the characteristic pre-peak due to mesoscopic structures in fluid-like systems.

No mesophase is observed in cases ii and iii. More precisely, for case ii, corresponding to a thin [C$_{12}$mim][Tf$_2$N] film 
adsorbed into the asymmetric potential well at a vacuum/solid interface, the optimal combination of Coulomb and dispersion interactions
is achieved by layering in the $z$ direction, without in-plane modulation. The alkane tails, in particular, tend to stick out of
the contact plane, leaving behind a thin layer of ionic moieties. In other terms, the film lowers its potential energy by exposing the inert alkane tails 
to the vacuum side, at the same time gaining entropy from the increased isomerisation freedom of tails in the peripheral location.
In-plane mesophases, therefore, may become stable only when a nearly 2D configuration is enforced by a sufficiently high normal load.
In case iii, Coulomb forces dominate the picture, giving rise to a fairly ordered in-plane ionic configuration. Excluded volume
and dispersion interactions are not sufficiently strong to bring tails together into neutral clusters, and thus the structure factor lacks the
pre-peak marking supra-molecular length scales. In other terms, similarly to what has been found for 3D systems, mesophases in 2D
are observed only for systems of sufficiently long tail, and the simulation results emphasise the need of comparable strength for Coulomb, packing
and dispersion forces. 

Temperature, 2D density and normal load ($\sim 10^2$ MPa) similar to those used in our simulations of case i
are found in RTIL's applications as lubricants\cite{lubricant}. Mesophases, therefore, could arise from 2D homogeneous systems through a phase transition, taking 
place when the applied normal load forces both tails (of sufficient length) and ionic heads to lay in a nearly 2D space of nanometric thickness.
The mesophase transition, in turn, could greatly affect the performance of the thin film as a lubricant.

Our investigation is different but related to a previous study\cite{voth}, devoted to the
computational investigation of surface manifestations of 3D mesophases in [C$_{n}$mim][X] RTIL's. Our results and those of
Ref.~\onlinecite{voth} are fully consistent with each other. At the free surface of RTIL's, ions find themselves in a local environment 
similar to ii, and in no cases they give rise to in-plane mesophases. Long-tail [C$_{n}$mim][X] systems, in particular, are not confined to a narrow
thickness by adsorption forces alone, and give rise to nanometric-thick layering perpendicular to the surface. Short-tail systems, instead, 
display monolayer ordering of alternating anions and cations, together with a random distribution of their alkane tails into the fluctuating 
interstices in between ions.

\section{The model and the computational method}
\label{method}
Molecular dynamics simulations have been carried out based on an AMBER\cite{amber} OPLS\cite{opls} all-atom potential. 
The potential energy as a function of atomic coordinates is the sum of intra- and inter-molecular terms. The
intra-molecular part includes stretching, bending and torsion contributions, while inter-molecular terms account for Coulomb and
dispersion interactions. These last two contributions are included also for atom pairs in the same connected unit ions) provided
they are separated by at least four covalent bonds. The parametrisation of Ref.~\onlinecite{canongia} (see also Ref.~\onlinecite{warn}) provides a comprehensive 
and thoroughly tested model of RTIL's of the alkane-imidazolium family, and we adopt this potential for our simulations of 
[C$_{12}$mim][Tf$_2$N] and [C$_{4}$mim][Tf$_2$N].

The systems we consider are nearly 2D, being extended along $xy$, and confined to nanometric thickness along $z$. 
However, Ewald sums are far more 
efficient for 3D periodicity than for quasi-2D systems. For this reason, we replicate periodically the simulation cell in 3D,
and account for long-range interactions by 3D Ewald sums. The convergence parameters for the direct- and reciprocal-space parts of the
Ewald sum have been set as follows: real space, screened Coulomb interactions are accounted for up to a separation $R_c^{Ew}=13.6$ \AA,
and reciprocal space terms are included whenever their weight $w(k)=\exp{[-k^2/4\alpha^2]}/k^2$ is larger than $10^{-5}$ \AA$^2$.
In the expression above, $\alpha=3.2/R_c^{Ew}$, and the ${\bf k}$'s are reciprocal lattice vectors of the periodically repeated simulation cell. 
The cut-off for the computation of dispersion interactions is set to $R_C^{vdW}=12$ \AA.
The sides of the simulation cell are kept fixed, and the periodicity $L_z$ along $z$ is much longer than either the actual width 
$\Delta z$ of the RTIL slab or the real space cut-offs $R_c^{Ew}$ and $R_c^{vdW}$, in such a way that we can neglect interactions 
among replicas in the $z$ direction.

\begin{figure}
\includegraphics[width=10cm]{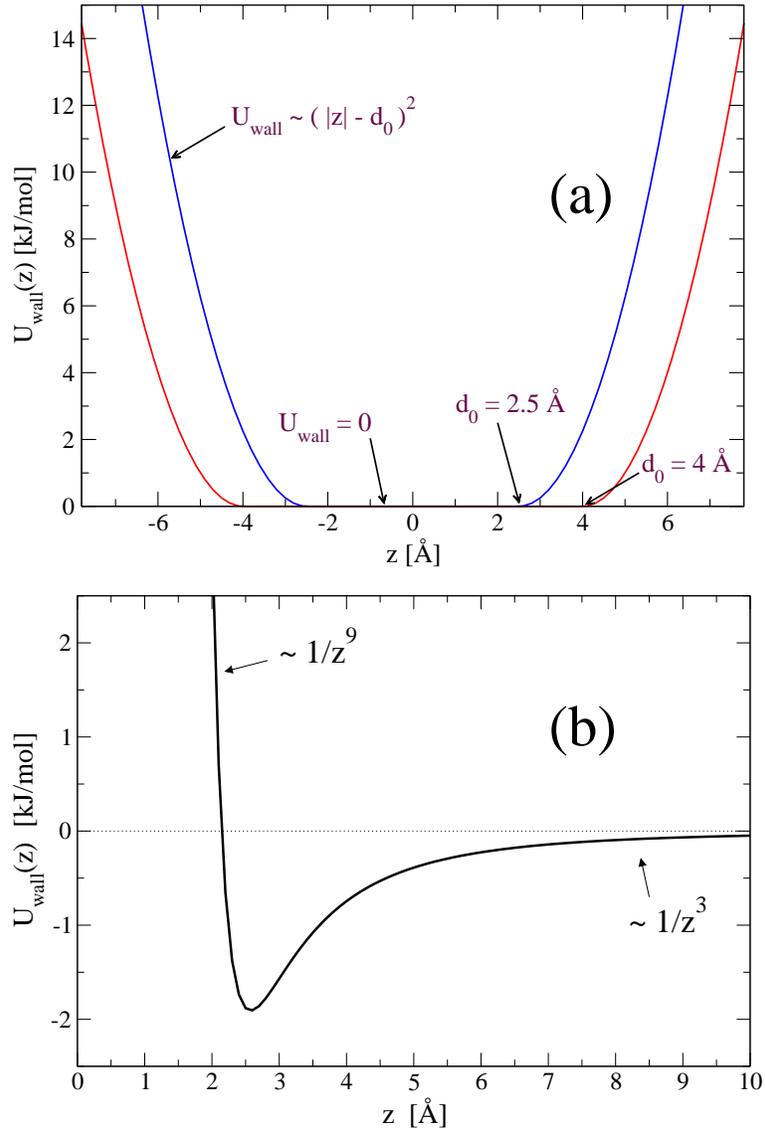}
\caption{Plot of the confining potentials used in the simulation. The potentials in (a) are symmetric, while the asymmetric
potential in panel (b) corresponds to the van der Waals potential outside an insulating solid surface.
}
\label{pote}
\end{figure}

The effect of confinement in between two planar surfaces perpendicular to the $z$ axis is represented by  a very
idealised, purely repulsive potential given by:
\begin{equation}
V(z)=\left\{
\begin{tabular}{cc}
$\frac{k_w}{2}(z+d)^2$ & $z < d$ \\
$0$ & $-d \leq z \leq d$ \\
$\frac{k_w}{2} (z-d)^2$ & $z > d$ \\
\end{tabular}
\right.
\label{wall1}
\end{equation}
which is applied to all RTIL atoms, see Fig.~\ref{pote}.
In all cases, the parameter $k_w$ is set to $k_w=1$ kJ/mol/\AA$^2 = 0.166$ N/m.
The parameter $d$, therefore, determines the separation of the two surfaces, and decreasing $d$ increases the
normal load applied to the interface. The actual thickness of the RTIL film exceeds $2d$, since atoms penetrate somewhat into the repulsive tail
of the potential, to an extent that depends on the system, on temperature,
and on pressure. For the systems and conditions we 
investigated, the film thickness can be estimated as $\Delta z\approx 2d+5$~\AA.
To explore the effect of different squeezing pressures, two values of $d$ ($d=2.5$~\AA\ and $d=4$~\AA) have been considered for the 
[C$_{12}$mim][Tf$_2$N] film, while keeping constant both the area of the interface and the number of RTIL molecules in the simulated sample.
Simulations for [C$_{4}$mim][Tf$_2$N] have been carried out only within the narrow ($d=2.5$~\AA) potential well.

As a comparison (see Fig.~\ref{pote}), we consider an asymmetric external potential meant to mimic the attractive well outside a solid surface
interacting with the adsorbate with dispersion forces:
\begin{equation}
V(z)=\frac{2 \pi \sigma_w^3 \epsilon_w \rho}{3} \left[ \frac{2}{15} 
\left(\frac{\sigma_w}{z}\right)^9-\left(\frac{\sigma_w}{z}\right)^3\right]
.
\label{wall2}
\end{equation}
The parameters $\sigma_w=3$~\AA, $\epsilon_w=0.8$ kJ/mol, and $4 \pi \rho=0.5$~\AA$^{-3}$ are gauged to reproduce the interaction 
of SiO$_2$ with carbon atoms \cite{nicholson} in a united-atoms representation of alkane chains.
As in the case of eq.~(\ref{wall1}), potential (\ref{wall2}) is applied to all RTIL atoms, and the model, therefore, does not
quantitatively represent any real interface. Nevertheless, we think that it offers the qualitative comparison that
it is intended to provide.

Molecular dynamics simulations have been performed using the DL$_\_$POLY package\cite{DLPOLY}, slightly modified 
to include the external potentials of the form (\ref{wall1}) and (\ref{wall2}).

To a large extent the analysis of the results relies on assigning the position of cations, anions and aliphatic chains. The position of the
two ions, in particular, is identified with the centre of charge of [C$_n$mim]$^+$ and [Tf$_2$N]$^-$, defined in full analogy with the centre of mass,
with atomic charges replacing masses in weighting the contribution of each atom. Since the alkane tail is neutral, and [Tf$_2$N]$^-$ is fairly symmetric,
this definition places the cation position at the centre of the imidazolium ring, and the anion position is close to the centre of mass of [Tf$_2$N]$^-$.
The location of aliphatic chains is identified with the position of their terminal carbon (CT carbon in Fig.~\ref{scheme}).

\section{Simulation results}
\label{results}
A system consisting of $N=256$ [C$_{12}$mim][Tf$_2$N] ion pairs  in the external potential of eq.~(\ref{wall1})
with $d=2.5$~\AA\ has been simulated by molecular dynamics in the NVT ensemble. The lateral periodicity in the 
$x$ and $y$ directions is set to $L_x=L_y=136$~\AA, corresponding to a
surface area $A = 185$~nm$^2$, and to
a surface density of $1.384$ ion pairs per nm$^2$.

To speed up equilibration, simulations have been carried out, first of all, at high temperature ($T=500$~K), then
$T$ has been progressively decreased in step of $50$~K. Relaxation, as detected by the drift in the running average of the 
potential energy following each temperature variation, turns out to be very slow, especially at the lowest temperatures.
In addition to the expected high viscosity of [C$_{12}$mim][Tf$_2$N], the long relaxation times are probably due to
the fairly high normal load applied to the simulated sample, increasing the activation barriers for all dynamical 
processes in the RTIL. At each temperature, the sample has been equilibrated for at least $200$~ps. Production runs 
lasting $10$ ns have been carried out at $T=500$~K, $T=400$~K, and $T=300$~K. 

The normal load applied to the RTIL film has been estimated by averaging the force on RTIL atoms due to the external potential. 
Since this potential depends only on  $z$, the applied force is
strictly directed along $z$. At constant area and number of ions, the measured load raises, as expected,
with increasing temperature, going from $P=276 \pm 0.5$~MPa at $T=300$~K, to $P=318 \pm 0.6$~MPa at $T=400$~K, and 
$P=359 \pm 0.7 $~MPa at $T=500$~K.
Pressures of this magnitude are fairly high but not exceptional for liquid-like films
in between approaching surfaces at microscopic separation. In fact, experiments on organic lubricant films of nanometric
thickness at ambient temperature have been extended up to pressures of several GPa\cite{press}. Moreover, diffusion coefficients
and electrical conductivity for several 3D RTIL samples have been reported in Ref.~\onlinecite{kanakubo} for pressures up to
$200$ MPa, and in Ref.~\cite{mahiuddin} up to $500$MPa (electrical conductivity only).

\begin{figure}
\includegraphics[width=10cm]{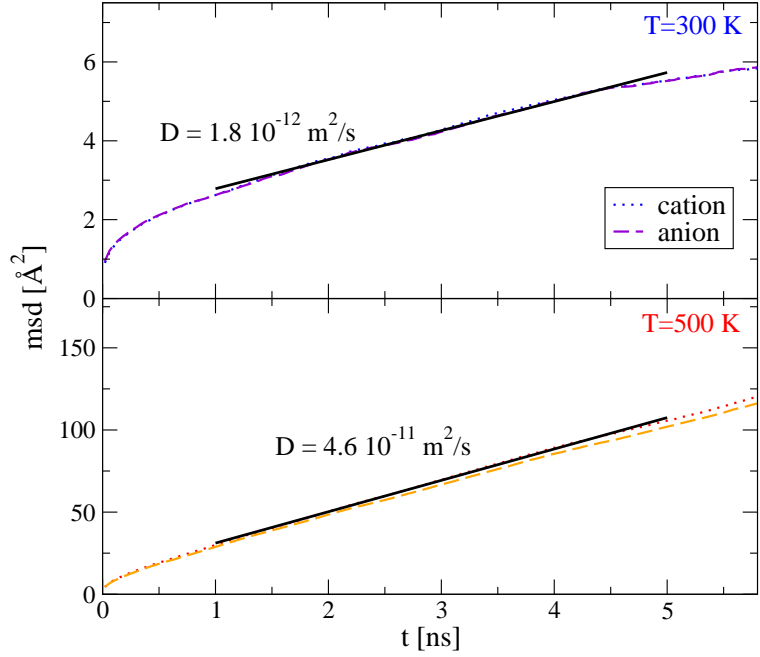}
\caption{Mean square displacement as a function of time for the centre of mass of cations ([C$_{12}$mim]$^+$) and anions ([Tf$_2$N]$^-$)
in the narrow symmetric confining potential ($d=2.5$~\AA). 
Note the change of scale from the upper ($T=300$~K) to the lower panel
($T=500$~K).
The solid straight lines are fit to the cation data in the $1 - 5$~ns range.
}
\label{msd}
\end{figure}

The film is liquid-like down to $T=300$~K, even though at this temperature and pressure the mobility is fairly low (see Fig.~\ref{msd}).
Ionic self-diffusion is quantified using the 2D version of Einstein's relation \cite{hmd}, i.e., by computing the mean square 
displacement of ions as a function of time. 
As already mentioned in Sec.~\ref{method}, we identify the position of ions with 
the centre of mass of the imidazolium ring (cation) and of [Tf$_2$N]$^-$ (anion). Linear fit of the mean square displacement 
for $1 \leq t \leq 5$ ns provides an estimate for the diffusion coefficient, that turns out to be surprisingly similar for 
anions and cations, already pointing to a strongly correlated motion for
the two species. The diffusion constant averaged over anions and cations is
$D=(4.6 \pm 0.2) \times 10^{-11}$~m$^2$/s at $T=500$~K,
$D=(7.0 \pm 0.5) \times 10^{-12}$~m$^2$/s at $T=400$~K,
and
$D=(1.8 \pm 0.2) \times 10^{-12}$~m$^2$/s at $T=300$~K. 
Comparison with experimental values of  self-diffusion coefficients measured on 3D samples is qualitative at best, mainly because of the different 
dimensionality (nearly 2D) of our systems. Moreover, it has been shown several times that computations using simple non-polarisable force fields tend
to under-estimate diffusion\cite{nonpol}. Nevertheless, self-diffusion coefficients of the order of $10^{-12}\div 10^{-11}$ m$^2$/s are 
routinely measured in experiments on imidazolium-based ionic liquids at ambient conditions\cite{diffex}, thus fully overlapping with the
range spanned by our results. Despite the uncertainties of the comparison, this observation is at least consistent with the conclusions of 
Ref.~\onlinecite{kanakubo}, showing that up to $\sim 200$ MPa, pressure reduces self-diffusion of imidazolium-based RTILs by at most one order of magnitude.


An Einstein-like relation has been used to estimate the ionic conductivity, based on the computation of the mean-square charge displacement:
\begin{equation}
\sigma=\frac{e^2}{K_BTA} \lim_{t\longrightarrow \infty} \frac{1}{4t} \langle \mid Z_+\Delta^+(t)+Z_- \Delta^-(t)\mid^2 \rangle 
\label{sig}
\end{equation}
where $\Delta^{\pm}=\sum_{i\in \pm} [{\bf R_i}(t)-{\bf R_i}(0)]$, $A$ is the area of the interface, and $Z_{+}$, $Z_-$ are equal to $+1$ and $-1$ for
cations and anions, respectively. In our computations, the limit implied in eq.~\ref{sig} is estimated from the average slope of
$\langle \mid Z_+\Delta^+(t)+Z_- \Delta^-(t)\mid^2 \rangle$ over the same time interval ($1 \leq t \leq 5$ ns) considered in the computation of the
diffusion coefficient.
The result, i.e., $\sigma=148\pm 6 \times 10^{-12} $ S at $T=500$ K, $\sigma=26.1 \pm  2 \times 10^{-12}$ S at $T=400$ K, 
and $\sigma=6.7\pm 1 \times 10^{-12}$ S at $T=300$ K deviates significantly from the prediction of the Nernst-Einstein relation:
\begin{equation}
\sigma_{NE}=\frac{n e^2}{K_B T} \left[ Z_+^2 D_++Z_-^2D_-\right]
\end{equation}
where $n$ is the 2D density of ion pairs, and $K_B$ is the Boltzmann constant. Introducing the values of the diffusion coefficient estimated by simulation
gives $\sigma_{NE}=484 \pm 20 \times 10^{-12} $ S at $T=500$ K, $\sigma_{NE}=90\pm 6 \times 10^{-12}$ S at $T=400$ K, and $\sigma_{NE}=31\pm 4.5
\times 10^{-12} $ S at $T=300$ K.
The deviation of molar conductivity from the Nernst-Einstein relation is often expressed by introducing the parameter\cite{harris} $\Delta$
defined as:
\begin{equation}
\sigma =\sigma_{NE} (1-\Delta)
\end{equation}
Then, for our system, we find $\Delta=0.78$ at $T=300$~K, $\Delta=0.71$ at $T=400$~K, and $\Delta=0.69$ at $T=500$~K, emphasising both the strong ionic
association in this system, and the slow thermal dissociation with increasing $T$. This is not surprising for [C$_{12}$mim][Tf$_2$N],
whose low average charge density favours bonding patterns somewhat different from those of ideal Coulomb systems.

Even more than in the case of self-diffusion, comparing the conductivity computed for our nearly 2D samples with experimental data is affected 
by the different dimensionality. As a result, the conductivities we obtain for our systems appear to be a few orders of magnitude lower than 
experimental values for the same or similar RTILs. We emphasise, however, that we are comparing quantities that are not even dimensionally equivalent.

\begin{figure}
\includegraphics[width=10cm]{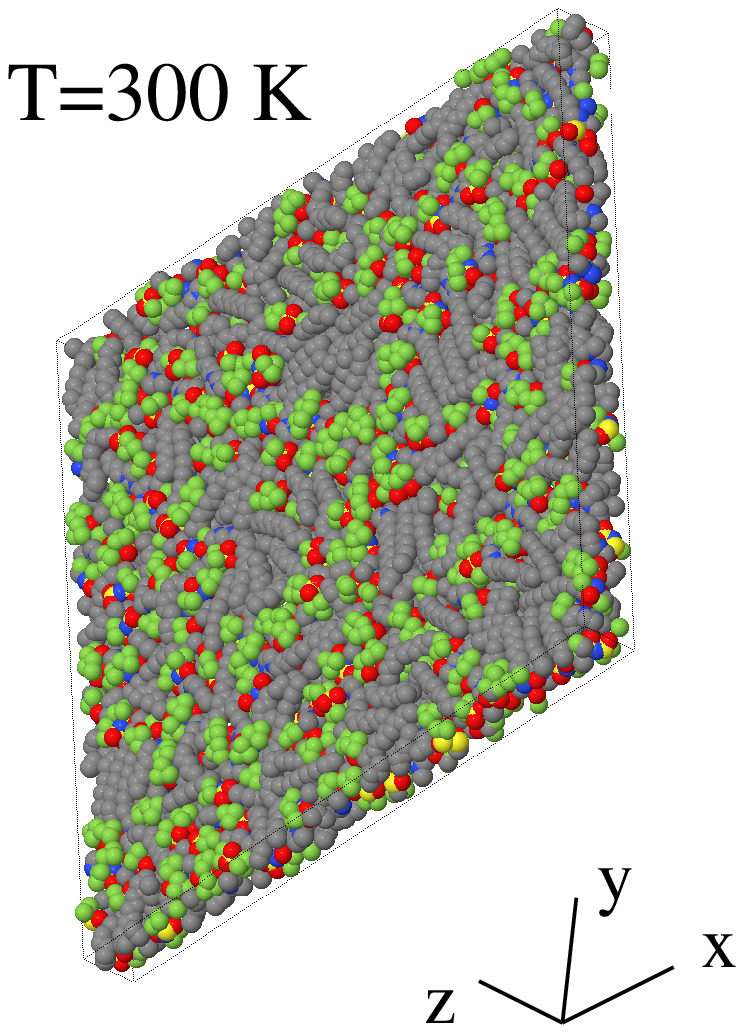}
\caption{Snapshot from the [C$_{12}$mim][Tf$_2$N] simulations in the $d=2.5$~\AA\ symmetric potential well.
The RTIL slab is tilted to show its aspect ratio.
}
\label{snap300}
\end{figure}

\begin{figure}
\includegraphics[width=10cm]{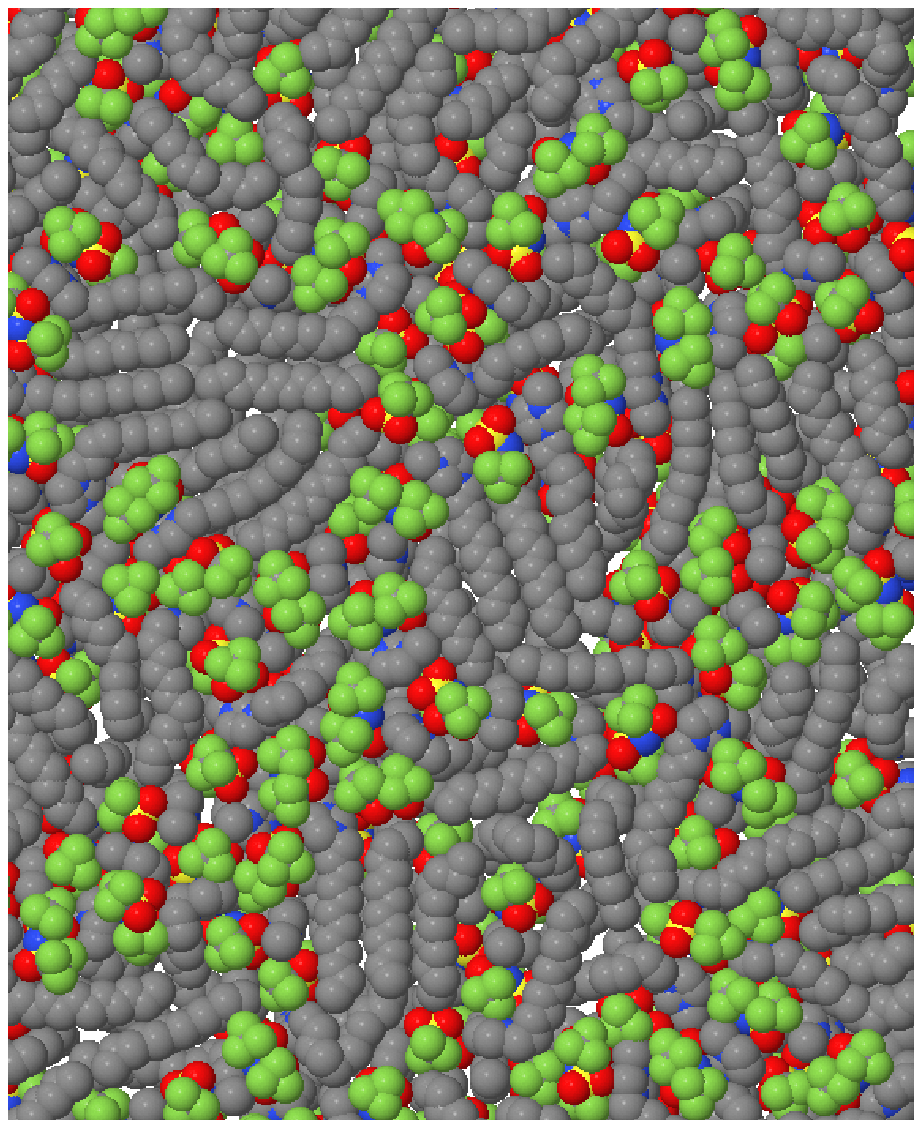}
\caption{Expanded view of the configuration of Fig.~\ref{snap300} seen from above. An area of $\sim 65 \times 80 $ \AA$^2$ is shown.
C: black particles; O: red; N: blue; S: yellow; F: green; Hydrogen atoms: not shown.
}
\label{snapzoom}
\end{figure}

\begin{figure}
\includegraphics[width=10cm]{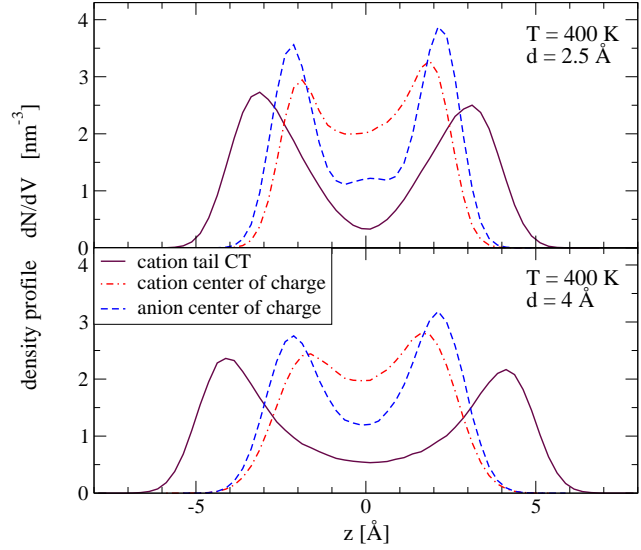}
\caption{Density distribution in the direction $z$ perpendicular to the interface for cations (red dash-dotted line), anions (blue dash line), and alkane tails 
(full line) in [C$_{12}$mim][Tf$_2$N] confined by the the narrow ($d=2.5$~\AA, upper panel) and wide ($d=4$~\AA, lower panel) symmetric potential well.
Anion and cation positions are identified by the centre of charge of the imidazolium ring and of [Tf$_2$N]$^-$, respectively.
The tail position is represented by the terminal CT carbon (see scheme in Fig.~\ref{scheme}).
}
\label{prof}
\end{figure}

A simulation snapshots from the [C$_{12}$mim][Tf$_2$N] simulation at $T=300$~K in the
$d=2.5$~\AA\ potential is displayed in Fig.~\ref{snap300}, while an expanded view of the same configuration is given in Fig.~\ref{snapzoom}.
These pictures already show that the applied load forces the imidazolium rings and the [Tf$_2$N]$^-$ anions to coexist in the central 
plane with the imidazolium alkane tails. These tails, in particular, lay parallel to the interfacial plane, overcoming energy and entropy forces that tend to align tails 
along $z$, and to separate them from the positively charged imidazolium rings and [Tf$_2$N]$^-$ anions.
A more quantitative analysis is based on the determination of the average orientation of the alkane tails (not shown), and on the
computation of the density and charge profiles (see Fig.~\ref{prof}).
This analysis confirms the validity of the qualitative picture arising from visual inspection of snapshots, 
even though it also reveals a slight separation of ionic moieties and neutral tails along the $z$ direction. 
The more ionic portions of [C$_{12}$mim][Tf$_2$N], in particular, tend to occupy the centre of the slab, 
while neutral tails tend to reside slightly outwards. The separation along $z$ seen in these computations, however, is not
comparable to the thick layering that has invariably been found in simulations of the RTIL's free surface\cite{voth, mario, milani}.
The effect of temperature on the density and charge profiles is marginal for $300 \le T \leq 500$.

\begin{figure}
\includegraphics[width=10cm]{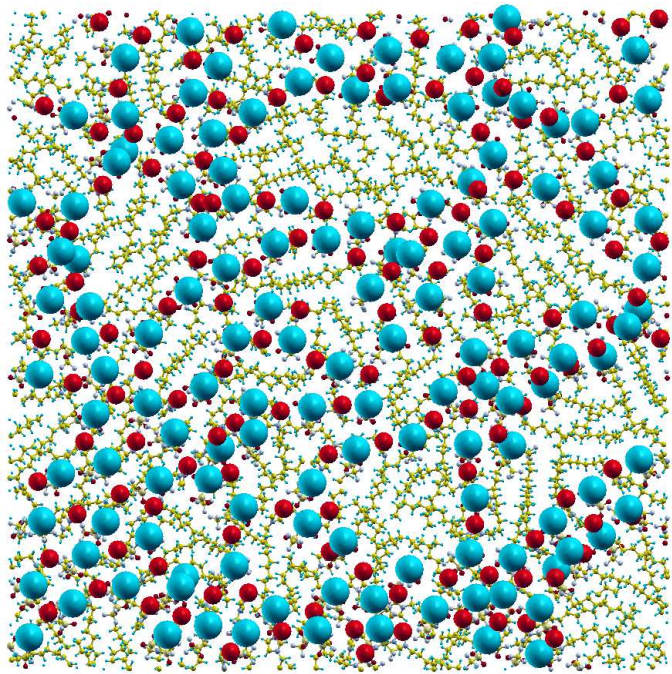}
\caption{Top view of a simulation snapshot for [C$_{12}$mim][Tf$_2$N] at $T=400$~K. Atoms are represented by small balls joined by covalent bonds using the following
colour coding: C: yellow particles;
O: red; N, F and S: brown; H: cyan. The large pale-blue balls represent the anion centre of charge, while the red balls of intermediate size
represent the centre of charge of the cation (see text).
}
\label{balls}
\end{figure}

Moreover, even a superficial analysis of simulation snapshots reveals a characteristic structure consisting of ion chains 
and hydrocarbon islands, which are apparent in the [C$_{12}$mim][Tf$_2$N] sample  at all simulated temperatures. These structures 
become better defined if we superimpose to each cation and each anion a single particle, located at the corresponding 
centres of charge (see Fig.~\ref{balls}). The system morphology highlighted in this way is clearly reminiscent of the modulated phases observed 
in 3D systems of analogous composition and temperature (see Ref.~\onlinecite{meso1, meso2, meso3, meso4}), collectively classified as mesophases. 
Systematic inspection of simulation snapshots already shows that the length scales of the density and composition modulation is of the order
of $2\div 3$ nm.

A few chain-like structures apparent in all simulation snapshots (see, for instance, Fig.~\ref{balls}), and made of the regular alternation of anions and cations,
point to important dipolar interactions in the system, confirming that [C$_{12}$mim][Tf$_2$N], as expected,
is far from behaving as an ideal Coulomb fluid. The tendency to ion association is probably increased also by the reduced system dimensionality, that
decreases the gain in Coulomb energy made available by dissociation.

\begin{figure}
\includegraphics[width=9cm,angle=-90]{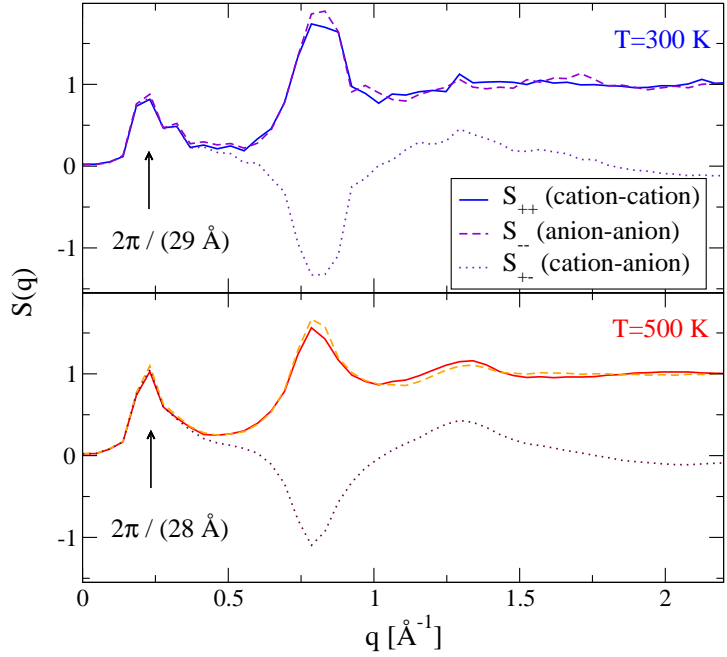}
\caption{Two-dimensional structure factors $S_{ij}(q)$ for anions and cations, whose positions are identified as in Fig.~\ref{balls}.
}
\label{sk}
\end{figure}

\begin{figure}
\includegraphics[width=9cm,angle=-90]{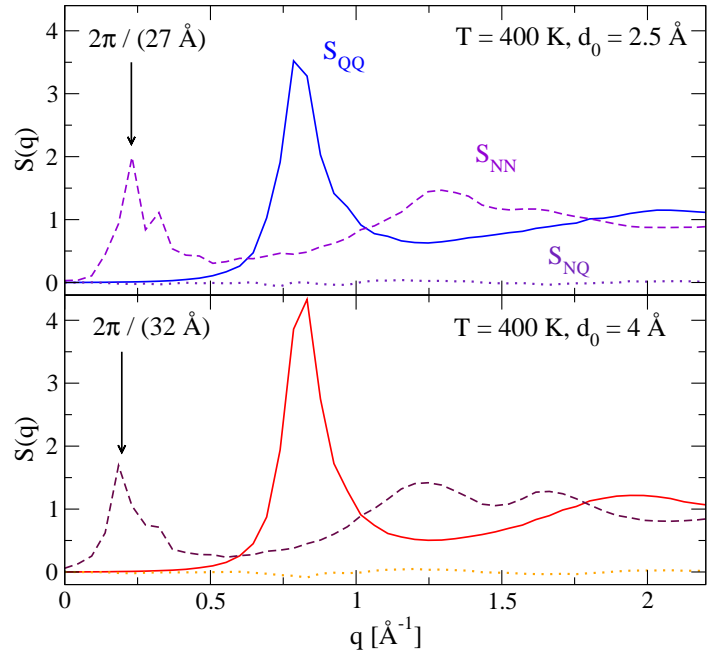}
\caption{Bathia-Thornton structure factors $S_{NN}$, $S_{QQ}$, $S_{NQ}$
  defined in ens.~(\ref{snn}-\ref{snq}) for anions and cations, whose
  positions are identified as in Fig.~\ref{balls}.
}
\label{skBT}
\end{figure}

We focus our analysis on the in-plane arrangement of ions, and we compute the in-plane (2D) ion-ion structure factors $S_{++}$, $S_{+-}$, $S_{--}$,
shown in Fig.~\ref{sk},
upon replacing again each cation and [Tf$_2$N]$^-$ anion by a single particle, as already done for visualising the
RTIL structure in Fig.~\ref{balls}. Even though this analysis is based directly only on the configuration of the ionic moieties,
the in-plane distribution of tails is described implicitly by the ionic structure factors, that provide a complementary 
view also of the alkane groups. The characteristic signature of mesophases is represented by a secondary peak in the structure factor
at a wave vector corresponding to a periodicity longer than the typical molecular size, that is reflected in the main peak of the
$S_{\alpha \beta}$. The mesophase formation, however, becomes more apparent if we combine the ionic structure factor into the 
Bathia-Thornton 
form\cite{hmd}:
\begin{eqnarray}\label{snn}
S_{NN}(q)&=&\frac{1}{2}[S_{++}(q)+S_{--}(q)+2S_{+-}(q)]
\\\label{sqq}
S_{QQ}(q)&=&\frac{1}{2}[S_{++}(q)+S_{--}(q)-2S_{+-}(q)]
\\\label{snq}
S_{NQ}(q)&=&\frac{1}{2}[S_{++}(q)-S_{--}(q)]
\,.
\end{eqnarray}
These combinations of the partial structure factors provide a fair diagonalisation of the $S_{\alpha \beta}(q)$ matrix, separating particle
particle (S$_{NN}$) from charge-charge ($S_{QQ}$) correlations, since the mixed charge-particle correlation nearly vanishes.
Moreover, and more importantly, as demonstrated in Fig.~\ref{skBT} a sizable pre-peak is
apparent in $S_{NN}$ at $q=0.225$~\AA$^{-1}$, supplementing the main peak corresponding to the average molecular size.
No pre-peak is apparent in the charge-charge partial structure factor,
probably because of the nearly equivalent size of the imidazolium ring and
[Tf$_2$N]$^-$ anion.

The pre-peak at $q\sim 0.225$~\AA$^{-1}$, points to a characteristic length of $28\div 29$~\AA\ for the mesophase structures.
Comparison of the results for $T=300$~K and $T=500$~K shows that the temperature dependence of the structure factors is fairly weak, and the
change in the pre-peak position and shape is comparable to the estimated error bar.

The mesoscopic islands consisting of alkane tails apparent in Fig.~\ref{balls} maintain their identity for times comparable to (and probably far exceeding) 
the entire duration of our simulations ($\sim 10$~ns), even though each island appears to diffuse slowly, as confirmed by computer animations 
generated from the simulation trajectories\cite{avail}.
No systematic drift in the characteristic size scale of the alkane tail islands
or of the ionic framework is observed beyond the equilibration stage, showing that no further aggregation or coalescence of
mesoscopic structures is taking place in the simulated sample. The structures shown in the snapshots and quantified by the
structure factors, therefore, appear to be equilibrium features of our systems.

The role of the applied normal load is analysed by performing simulations for a system made of the same
number of ion pairs placed into a wider potential well ($d=4$~\AA). Simulations have been carried out at $T=400$~K only, starting 
from the system equilibrated in the narrow slit at the same temperature. The [C$_{12}$mim][Tf$_2$N] film has been re-equilibrated in the new potential 
for $200$~ps. As expected, mobility in the $d=4$~\AA\ slit is higher ($D=1.9\pm 0.2 \times 10^{-11}$ m$^2$/s, 
averaged over cations and anions) than in the $d=2.5$~\AA\ case, enhancing equilibration that, however, at $T=400$~K seems to be easy in 
both cases. Also in this case production runs lasted $10$~ns.
The normal load turns out to be $105.4 \pm 0.2 $~MPa, to be compared to 
$318 \pm 0.6$~MPa in the narrower potential. Despite the sizable difference in the applied normal load, the system morphology is fairly similar 
in the $d=2.5$~\AA\  and in the $d=4$~\AA\ systems. Alkane tails, in particular, still lay close to the $z=0$ plane even within the $d=4$~\AA\ potential.
The density profiles for ions and tails, again as expected, are broader than in
the $d=2.5$~\AA\ case, but otherwise are qualitatively similar for the two
potentials (see Fig.~\ref{prof}).
Separation of tails and ions along $z$ is slightly enhanced in going from $d=2.5$~\AA\ to $d=4$~\AA, but neutral and ionic groups are still
closely intermixed. Nearly 2D alkane tail islands are still apparent in the wider potential, and their characteristic length scale 
is similar in the two cases. The similarity is confirmed by the computation of the in-plane structure factors, shown in Fig.~\ref{skBT}.
The differences between the $S(q)$'s for the two potentials are not much larger than the estimated error bar, especially for what concerns the pre-peak 
at $q\sim 0.225$~\AA$^{-1}$. Qualitative differences, however, are expected to take place with further decreasing the normal load, when layering of
ions and neutral tails along $z$ will start to be energetically admissible.

A further test of the system properties at even lower normal load is provided by the simulation of [C$_{12}$mim][Tf$_2$N] in the asymmetric
potential well of eq.~(\ref{wall2}).
This case represents, in fact, an idealised picture of a RTIL nanometric film physisorbed onto a solid and de-hydroxylated SiO$_2$ surface.
The major difference with the case of eq.~(\ref{wall1}) is that, in the case of eq.~(\ref{wall2}), the  potential well has finite height
($\Delta U=1.9$~kJ/mol, or $\Delta U=19.7$~meV {\it per atom})
and finite slope on the $z \longrightarrow \infty$ side, thus limiting the normal load that can be applied on the trapped RTIL molecules
by the adsorption potential.
The finite value of $\Delta U$ already suggests that the RTIL can form a compact droplet instead of an homogeneous film, provided the reduction in 
surface area, and thus the gain in surface free energy, overcomes the (limited) loss of adsorption energy. 
The de-wetting of the film, therefore, will depend on the relative size of the surface and interfacial tension compared to the adsorption energy. 
In the case of [C$_{12}$mim][Tf$_2$N] in the model potential of eq.~(\ref{wall2}), the surface free energy apparently is the dominant term, and 
simulations continued for $10$~ns after the initial equilibration show the beginning of a clear de-wetting. Already at this stage, a sizable fraction of 
the RTIL molecules sit outside the central part of the adsorption potential. Any estimate of the normal load on the ionic liquid due to adsorption, therefore,
does not reflect the internal state of ions over the entire sample, and it is not comparable to the values computed in part i.
The RTIL configuration at the end of these simulations is reminiscent of the structures already seen in computer simulations of free surfaces and 
interfaces of room temperature ionic liquids. Aliphatic tails orient themselves toward the vacuum side, giving rise to
relatively dense outermost layer of nanometric thickness. Exposing the neutral tails to the vacuum side decreases the
surface tension of the interface down to values ($\gamma\sim 0.05$ N/m, see Ref.~\onlinecite{delamora}) typical of alkane hydrocarbon liquids,
and much lower than those of inorganic ionic liquids. For what concerns our discussion, however, the important point is that, as already observed, 
the optimal separation of ions and neutral tails occurs along the direction perpendicular to the interface, pre-empting the formation of mesophases, 
whose potential energy advantage is partly compensated by a decrease of ideal (translational) entropy in forming the mesoscopic islands.

Since the [C$_{12}$mim][Tf$_2$N] system considered in our study is a fairly typical RTIL, it is expected that the system evolution shown by our 
simulations for case ii, is the relevant one for many RTIL films deposited on insulating and neutral solid interfaces. Variations with respect to this
picture, however, might be expected in the case of specific interactions of RTIL's  with the surface, due, for instance, to the formation of hydrogen bonds.
For instance, preliminary simulation and experimental results for [C$_4$mim][Tf$_2$N] films on hydroxylated SiO$_2$ surfaces have been reported in Ref.~\onlinecite{milani},
showing clear layering both at the solid/RTIL and RTIL/vacuum interfaces, no in-plane mesophase, and no de-wetting, very likely because of the high 2D density
of hydrogen bonds connecting RTIL ions to the OH groups at the silica surface.

\begin{figure}
\includegraphics[width=10cm]{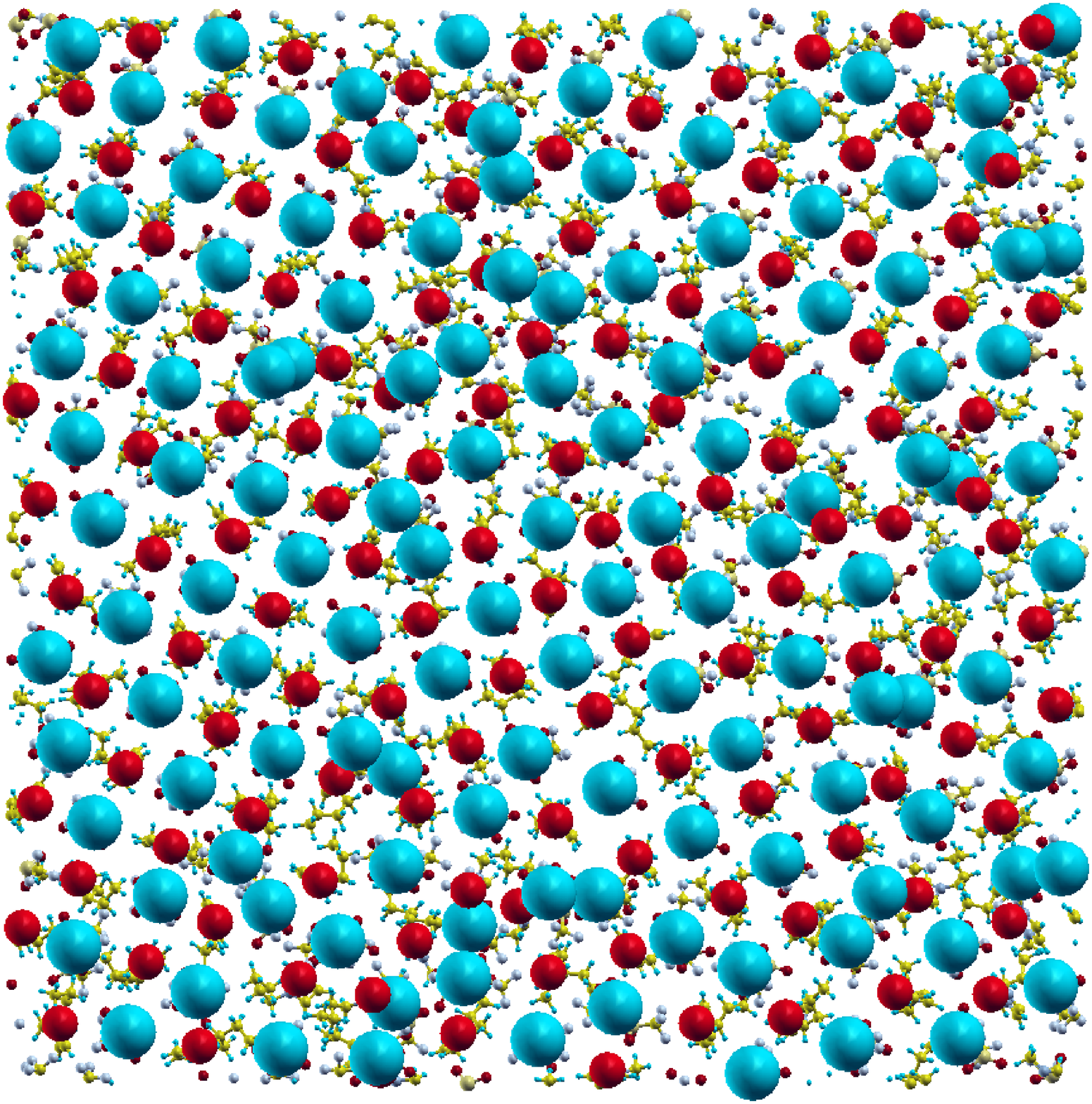}
\caption{Snapshot from the simulation trajectory of [C$_4$mim][Tf$_2$N] in the narrow ($d=2.5$ \AA\ ) symmetric potential at $T=400$ K.
The same colour coding of Fig.~\ref{balls} has been used.
Imidazolium rings and [Tf$_2$N]$^-$ anions, in particular, are represented by red and blue particles, respectively, located at their centre of charge.
}
\label{scSnap}
\end{figure}

Finally, the results for case iii differ from those of both i and
ii. Computations for [C$_4$mim][Tf$_2$N]
in the narrow symmetric
potential have been carried out at $T=400$~K with $N=128$ ion pairs
in a simulation cell of reduced area ($A= 78 \times 78$~\AA$^2$) in the periodic plane, in such a way to
reproduce a normal load close to that of case i: in case iii we obtain
$528 \pm 1.0$~MPa.
Reducing the size of the aliphatic chain increases the average charge density, and thus the Coulombic character of the system.
This is reflected into a fairly high degree of order, that can be appreciated in Fig.~\ref{scSnap}. Ions of alternating sign arrange themselves into 
crystal-like grains of nanometric size, with alkane tails occupying the interstices. A sizable population of point and extended defects prevents the onset of
long range order, which, on the other hand, is the defining property of real crystals. As a result, the system appears to be liquid, or at least glassy,
as confirmed by the computation of the structure factors, that are clearly fluid-like. On the time scale of our simulations, the diffusion coefficient 
$D=(4.1 \pm 0.9) \times 10^{-12}$~m$^2$/s of [C$_4$mim][Tf$_2$N] at $T=400$~K is not negligible, but, despite a sizable mass advantage, it is even 
smaller than that of [C$_{12}$mim][Tf$_2$N] at the same thermodynamic conditions. No pre-peak appears in any of the [C$_4$mim][Tf$_2$N] structure factors.

The result for the iii simulations, therefore, confirm that neutral tails of sufficient length are required to give rise to mesophases and to the 
pre-peaks that are their distinctive diagnostic feature.

The absence of long range order in a system that locally tends to adopt an ionic crystal-like structure is likely to be due, first of all,
to the limited time allowed by simulation, that is typically shorter than the time needed to reach a well ordered configuration, even when this is
the most stable one from a thermodynamic point of view.
However, it might also be due to the random perturbation represented by the floppy neutral tails, which might prevent long range order.

\section{Summary and conclusions}

Molecular dynamics simulations for a thin film made of a long-chain ionic liquid ([C$_{12}$mim][Tf$_2$N]) trapped in between planar and rigid surfaces at kbar pressures
reveal stable in-plane mesoscopic structures (mesophases) consisting of ion clusters and of neutral islands, formed by the alkane tails carried by the cation.

Computations for [C$_4$mim][Tf$_2$N] in the same potential, developing no mesophases at comparable thermodynamic conditions,
show that aliphatic chains of sufficient length are needed to form nanometric islands, bringing together tails from different cations.

The mesoscopic structures found in our nearly 2D systems are qualitatively similar to those seen in 3D simulations of similarly sized RTIL's
at comparable thermodynamic conditions. Then, comparison with the results for the shorter-chain [C$_4$mim][Tf$_2$N] case suggests that
the molecular mechanism responsible for mesophases is similar in two and in three dimensions, and relies on the competition
of strong Coulomb forces with weaker but nevertheless comparable dispersion and steric interactions.

Our simulations show that strict confinement in a narrow slit is an essential requirement for the formation of mesophases.
As soon as molecules can vary their orientation, like, for instance, in the weaker confining potential of case ii, the optimal reciprocal
arrangement of ionic and neutral groups is achieved by layering in the direction $z$ perpendicular to the surface, following a trend
already apparent in free surfaces\cite{voth, mario, milani}. The system preference for this configuration is based on a few different effects:
exposing the aliphatic tails to the vacuum side or to heterogeneous phases lowers the surface/interfacial tension to values
($\sim 0.05$~N/m, see Ref.~\onlinecite{delamora}) typical
of liquid hydrocarbons, and lower than those of inorganic ionic liquids; tails gain entropy from the orientational freedom at the interface;
the nanometric separation of charges and neutral tails, while decreasing entropy, optimises the packing of charges and of neutral tails,
gaining both Coulomb and dispersion energy, while decreasing short-range repulsive contacts.

All these observations together suggest that mesophases could be relevant for RTIL's used as lubricants in between solid surfaces, when
local temperatures and normal load become comparable to those of our simulations. Mesophases, in particular, could form in RTIL of sufficiently
long tail with decreasing separation of the lubricated surfaces, when the applied load reaches values in the kbar range, entering the regime
of boundary lubrication  and squeezing the RTIL into a nearly 2D space.

The formation of mesophases could affect technologically important properties such as viscosity, and could be verified by
measurements using a diamond anvil cell\cite{anvil}.

Besides these reason of applied interest, the formation of mesophases in nearly-2D RTIL systems, if confirmed by experiments and by further computations,
would add another fascinating chapter to the already rich phenomenology of these low temperature Coulombic fluids.

ACKNOWLEDGEMENTS - Computing resources were kindly provided by the Consorzio Interuniversitario Lombardo per l'Elaborazione Automatica (CILEA). 

\begin{thebibliography}{99}
\bibitem{wasser} Wasserscheid, P.; Welton, T. {\it Ionic Liquids in Synthesis}; 
Wiley-VCH, Weinheim, 2002.
\bibitem{welton} Welton, T. {\it Chem. Rev.} {\bf 1999}, {\it 99}, 2071;
Blanchard, A.; Hancu, D.; Beckmann, E. J.; Brennecke, J. {\it Nature} {\bf 1999}, {\it 399}, 28.
\bibitem{green} {\it Ionic liquids: Industrial applications for
green chemistry} edited by R. D. Rogers and K. R. Seddon, (American 
Chemical Society, Washington DC, 2002).
\bibitem{lubricant} Fox, M. F.; Priest, M. {\it Proc. IMechE} {\bf 2008}, {\it 222 J}, 291;
Ye, C.-F.; Liu, W. M.; Chen Y. X.; Yu, L. G. {\it Chem. Commun.} {\bf 2001}, 2244–2245;
Wang, H. Z.; Lu, Q. M.; Ye, C. F.; Liu, W. M.; Cui, Z. J. {\it Wear} {\bf 2004}, {\it 256}, 44.
Mu, Z.-G.; Liu, W. M.;  Zhang, S. X.; Zhou, F. {\it Chem. Lett.} {\bf 2004}, {\it 33} 524.
\bibitem{popolo} Del P{\'o}polo, M. G.; Mullan, C. L.; Holbrey, J. D.; Hardacre, C.; Ballone, P.
{\it J. Am. Chem. Soc.} {\bf 2008}, {\it 130},  7032-7041.
\bibitem{meso1} Urahata, S. M.; Ribeiro, M. C. C. {\it J. Chem. Phys.} {\bf 2004}, {\it 120}, 1855-1863.
\bibitem{meso2} Wang, Y.; Voth, G. A. {\it J. Am. Chem. Soc.} {\bf 2005}, {\it 127}, 12192.
\bibitem{meso3} Canongia Lopes, J. N. A.; Padua, A. A. H. {\it J. Phys. Chem. B} {\bf 2006},
{\it 110}, 3330.
\bibitem{meso4} Wang, Y.; Voth, G. A. {\it J. Phys. Chem. B} {\bf 2006}, {\it 110}, 18601.
\bibitem{expemeso} Triolo, A.; Russina, O.; Bleif, H.-J.; Di Cola, E. {\it J. Phys. Chem. B} {\bf 2007}, {\it 111}, 4641.
\bibitem{voth} Jiang, W.; Wang, Y.; Yan, T.; Voth, G.A.  {\it J. Phys. Chem. C} {\bf 2008}, {\it 112}, 1132.
\bibitem{amber} Cornell, W. D.; Cieplak, P.; Bayly, C. L.; Gould, I. R.; Merz, K.M. Jr.; 
Ferguson, D. M.; Spellmeyer, D. C.; Fox, T.; Caldwell, J. W.;
Kollman, P. A. {\it J. Am. Chem. Soc. } {\bf 1995}, {\it 117}, 5179.
\bibitem{opls} Jorgensen, W. L.; Maxwell, D. S.; Tirado-Rives, J. {\it J. Am. Chem. Soc.} {\bf 1996},
{\it 118}, 11225. 
\bibitem{canongia} J. N. Canongia Lopes, J. Deschamps, and A. A. H. Padua, 
{\it J. Phys. Chem. B} {\bf 2004}, {\it 108}, 2038; Canongia Lopes, J. N.; Padua, A. A. H.
{\it J. Phys. Chem. B} {\bf 2004}, {\it 108}, 16893. See also (Additions and corrections): 
Canongia Lopes, J. N.; Deschamps, J.; Padua, A. A. H. {\it J. Phys. Chem. B } {\bf 2004},
{\it 108}, 11250.
\bibitem{warn} It is important to note that the mixing rules for the
Lennard-Jones coefficients adopted in Ref.~\onlinecite{canongia} are: 
$\sigma_{ij}=\sqrt{\sigma_{ii} \sigma_{jj}}$,
$\epsilon_{ij}=\sqrt{\epsilon_{ii}\epsilon_{jj}}$, and differ from the popular Berthelot 
rules. Canongia Lopes, J. N.; Padua, A. A. H. {\it private communication}.
\bibitem{nicholson} Nicholson, D.; Parsonage, N. G., {\it Computer Simulation and the
Statistical Mechanics of Adsorption}; Academic Press: London, 1982.
\bibitem{DLPOLY} Smith, W.; Leslie, M.; Forester, T. R. DL$_\_$POLY v2.14;
Daresbury Laboratories: Daresbury, Warrington, WA4 4AD, UK, 2003.
\bibitem{press} Spikes, H. A. {\it Proc. Instn. Mech. Engrs.} {\bf 1999}, {\it 213J}, 335; Bair, S.;
Winer, W. O. {\it J. Tribology} {\bf 1992}, {\it 114}.
\bibitem{kanakubo} Harris, K. R.; Kanakubo, M.; Tsuchihashi, N.; Ibuki, K.; Ueno, M. {\it J. Phys. Chem. B} {\bf 2008},
{\it 112}, 9830.
\bibitem{mahiuddin} Mahiuddin, S.; Rohman, N.; Aich, R.; T{\"o}dheide, K. {\it Aust. J. Chem.} {\bf 1999}, {\it 52}, 373. 
\bibitem{nonpol} Kelkar, M. S.; Shi, W.; Maginn, E. J. {\it Ind. Eng. Chem. Res. } {\bf 2008}, {\it 47}, 9115;
Bagno, A.; D'Amico, F.; Saielli, G. {J. Mol. Liq.} {\bf 2007}, {\it 131-132}, 17.
\bibitem{diffex} See, for instance, Umecky, T.; Kanakubo, M.; Ikushima, Y. {\it Fluid Phase Equilib.} {\bf 2005}, {\it 228-229}, 329;
Annat, G.; MacFarlane, D. R.; Forsyth, M. {\it J. Phys. Chem. B} {\bf 2007}, {\it 111}, 9018.
\bibitem{hmd} Hansen, J.-P.; McDonald, I. R. {\it Theory of Simple Liquids}; Academic
Press: San Diego, CA, 1996.
\bibitem{sigma2D} The quantity $\sigma$ defined in eq.~\ref{sig} is a 2D conductivity, and, has such, is measured in Siemens (SI units), 
instead of Siemens/m as it would be the case for a 3D system.
\bibitem{harris} See the introductory pages of Ref.~\onlinecite{kanakubo}.
\bibitem{mario} Lynden-Bell R. M., Del Popolo M. G. {\it Phys. Chem. Chem. Phys.} {\bf 2006}, {\it 8}, 949-954.
\bibitem{milani} Bovio, S.; Podest{\'a}, A.; Milani, P.; Ballone, P.; Del P{\'o}polo, M. G. {\it J. Phys.: Condens. Matter} 
{\bf 2009}, {\it 21}, 424118.
\bibitem{avail} Two sample movies are provided as supplementary material.
Additional trajectories and movies are available upon request from the authors.
\bibitem{delamora} Martino, W.; de la Mora, J. F.; Yoshida, Y.; Saito, G.; Wilkes, J.
{\it Green Chemistry} {\bf 2006}, {\it 8}, 390-397. 
\bibitem{anvil} Presser, V.; Krummhauer, O.; Nickel, K. G.; Kailer, A.; Berthold, C.; Raish, C. {\it Wear} {\bf 2009}, {\it 266} 771-781;
Piermarini, G. J.; Forman, R. A.; Block, S. {\it Rev. Sci. Instrum.} {\bf 1978}, {\it 49}, 1061-1066.
\end{thebibliography}
\end{document}